\documentclass[twocolumn,showpacs,preprintnumbers,amsmath,amssymb]{revtex4}

\usepackage{graphicx}% Include figure files                                     
\usepackage{dcolumn}% Align table columns on decimal point                      
\usepackage{bm}% bold math                                                      

\usepackage{latexsym}
\usepackage{amssymb}
\usepackage{amsbsy}
\usepackage{amsmath}
\usepackage{epsfig}
\usepackage{graphics,color}

\newcommand{\be}{\begin{equation}}
\newcommand{\ee}{\end{equation}}
\newcommand{\bea}{\begin{eqnarray}}
\newcommand{\eea}{\end{eqnarray}}
\newcommand{\bean}{\begin{eqnarray*}}
\newcommand{\eean}{\end{eqnarray*}}

\definecolor{DarkGreen}{rgb}{0,0.5,0}

\begin{document}

%}}}

%{{{ title/abstract

\title{Ab Initio Calculation of Finite Temperature Charmonium Potentials}

\author{P.W.M.~Evans${}^a$, C.R.~Allton${}^a$ and J.-I.~Skullerud${}^b$
  \\
{${}^a$\em\normalsize Department of Physics, Swansea University,
  Swansea, United Kingdom} \\
{${}^b$\em\normalsize Department of Mathematical Physics, National
  University of Ireland Maynooth} \\
{\em\normalsize Maynooth, County Kildare, Ireland}
}

\date{\today}

\begin{abstract}
The interquark potential in charmonium states is calculated for the
first time in both the zero and non-zero temperature phases from a
first-principles lattice QCD calculation. Simulations with two
dynamical quark flavours were used with temperatures $T$
in the range $0.4 T_c \lesssim T \lesssim 1.7 T_c$, where $T_c$ is the
deconfining temperature. The correlators of point-split operators were
analysed to gain spatial information about the charmonium states.  A
method, introduced by the HAL QCD collaboration and based on the
Schr\"odinger equation, was applied to obtain the interquark
potential. We find a clear temperature dependence, with the central
potential becoming flatter (more screened) as the temperature
increases.
\end{abstract}

 % Heavy quarkonia

\maketitle

%}}}

%{{{ Introduction

{\em Introduction --}
The quark-gluon plasma (QGP) phase of QCD has been studied extensively
both in heavy-ion collision experiments at RHIC
\cite{Adams:2005dq,Adcox:2004mh} and the LHC \cite{Aamodt:2009aa} as
well as in theoretical calculations. However, a complete understanding
of this phase is still some distance away.  Experiments are hindered
by uncertainties in the phenomenology of the QGP such as the equation
of state, transport properties, and spectral features of
hadrons. These quantities are required to model the QGP fireball in
heavy-ion collisions as it expands and cools back into the hadronic
phase in order that the events in the detectors can be properly
interpretted.

One of the quantities of interest is the interquark potential in the
QGP phase.  A temperature dependent charmonium potential underlies the
widely cited J/$\psi$ suppression model of Ref.
\cite{Matsui:1986dk}. More recent work on statistical models of
charmonium production
\cite{BraunMunzinger:2000px,BraunMunzinger:2000ep} and studies
assuming transport models of charmonium production
\cite{Liu:2009nb,Zhao:2011cv} lead to alternative interpretations.  An
analogous suppression has recently been found in bottomonium yields in
heavy-ion collisions \cite{Chatrchyan:2011pe,Reed:2011fr}.

Theoretical work on the interquark potential at high
  temperature includes early models
\cite{Karsch:1987pv} and perturbative QCD calculations
\cite{theory-pot}.  Furthermore, there have been some recent
non-perturbative (i.e. lattice) QCD studies of interquark potentials
which are relevant to the work presented here. These fall into two
categories: (i) non-zero temperature studies of the {\em static} quark
potential
\cite{static-pot,Kaczmarek:2005ui,Bazavov:2012fk,rothkopf,Burnier:2012az}
and (ii) {\em zero temperature} studies of the potential between
quarks with finite masses \cite{nbs}. The work presented here is a
study of the interquark potential of charmonium using {\em physical
  charm quark masses} at {\em finite temperature} and uses
two flavours of light dynamical quark. A particular feature of our
work is that our lattices are anisotropic which has the significant
advantage that our correlation functions are determined at a
large number of temporal points, hence aiding our analysis.

The method we use is based on the HAL QCD collaboration's calculation
of the {\em internucleon} potential relevant for nuclear physics and
utilised the Schr\"odinger equation \cite{halqcd}. In this work we use
their ``time-dependent'' method \cite{time-dep} to determine the real
part of the {\em interquark} charmonium potential.  In our work we do
not consider the width of the state and therefore have access to the
real part of the potential only.  The possible limitations of
the underlying assumption, that a nonrelativistic potential
description is valid for these temperatures and quark masses, is a
separate issue which will not be discussed here.

Our main conclusion is that the charmonium potential as a
function of distance is steepest for low temperatures $T$, and
becomes flatter at large distances as $T$ increases. This work extends our
earlier work in \cite{Allton:2012ki}.

%}}}

%{{{ Time-dependent Schr\"odinger Equation Approach

{\em Time-dependent Schr\"odinger Equation Approach --}
Following HAL QCD, we determine the potential
using their ``time-dependent'' method \cite{time-dep}.
The first step is to define charmonium point-split operators,
\begin{equation}
J_\Gamma(x;\mathbf{r}) = q(x) \,\Gamma\, U(x,x+\mathbf{r}) \, \overline{q}(x+\mathbf{r}),
\end{equation}
where $\mathbf{r}$ is the displacement\footnote{Only on-axis separations were
  studied in this work.} between the charm and anti-charm quark fields
$q$ and $\overline{q}$, $x$ is the space-time point
$(\mathbf{x},\tau)$ and $\Gamma$ is a
Dirac matrix used to generate vector ($J/\psi$) or pseudoscalar
($\eta_c$) channels.  $U(x,x\!+\!\mathbf{r})$ is the gauge connection between
$x$ and $x+\mathbf{r}$.  The correlation functions,
\begin{equation}
C_\Gamma(\mathbf{r},\tau) =
\sum_{\mathbf{x}} \langle J_\Gamma(\mathbf{x},\tau;\mathbf{r}) \; J_\Gamma^\dagger(0;\mathbf{0}) \rangle.
\end{equation}
of the point-split and local operators, can be expressed in the usual
spectral representation,
\begin{equation}
C_\Gamma(\mathbf{r},\tau) = \sum_j \frac{\psi_j^\ast(\mathbf{0}) \psi_j(\mathbf{r})}{2E_j}
\;\left( e^{-E_j \tau} + e^{-E_j (N_\tau - \tau)} \right),
\label{eq:cfn}
\end{equation}
where the sum is over the states $j$ with the same quantum numbers as
the operator $J_\Gamma$, and $\psi_j(\mathbf{r})$ are the
corresponding Nambu Bethe Salpeter (NBS) wavefunctions. $N_\tau$ is the
number of lattice points in the temporal direction and is related to
the temperature by $T = 1/(a_\tau N_\tau)$, where $a_\tau$ is the
temporal lattice spacing.

From now on we consider only radially symmetric (S-wave) states.
We differentiate Eq.(\ref{eq:cfn}) w.r.t. time and apply the
Schr\"odinger equation which, in Euclidean space-time is
\begin{equation}
\left[- \frac{1}{2\mu} \frac{\partial^2}{\partial r^2} + V_\Gamma(r)
  \right] \psi_j(r)
= E_j \psi(r),
\end{equation}
where $\mu$ is the reduced mass of the $c\bar{c}$ system,
$\mu = \frac{1}{2} m_c \simeq \frac{1}{4} M_{J/\psi}$.
Ignoring the backward moving contribution, we obtain
\begin{align} 
\frac{\partial C_\Gamma(r,\tau)}{\partial \tau}
&= \!\sum_j
\left( \frac{1}{2\mu} \frac{\partial^2}{\partial r^2}
             - V_\Gamma(r) \right)
\frac{\psi_j^\ast(0) \psi_j(r)}{2E_j} e^{-E_j \tau} \notag\\
&=
\left( \frac{1}{2\mu} \frac{\partial^2}{\partial r^2}
             - V_\Gamma(r) \right) C_\Gamma(r,\tau).
\label{eq:V}
\end{align}
This can be trivially solved for the potential $V_\Gamma(r)$.

Notice that the NBS wavefunction, $\psi(r)$, is not explicitly
required in the above derivation of $V(r)$.  However, we note that HAL
QCD's original ``wavefunction'' method extracts $\psi(r)$ from a fit
to the large time behaviour of the correlation function, $C(r,\tau)
\rightarrow \psi_0(0) \psi_0(r) \; e^{-E_j \tau},$ and then uses this
$\psi_0(r)$ as input into the Schr\"odinger equation to obtain the
potential \cite{nbs}. HAL QCD's time-dependent method used here has
the distinct advantage that the correlation functions are used
directly, without requiring a fit to the asymptotic state.

The S-wave potential can be expressed as
\begin{equation}
V_\Gamma(r) = V_C(r) + \mathbf{s}_1\cdot \mathbf{s}_2 \;V_S(r),
\label{eq:pot-cs}
\end{equation}
where $V_C$ is the spin-independent (or ``central'') potential,
$V_S$ is the spin-dependent potential, and $\mathbf{s}_{1,2}$ are the
spins of the quarks. We have $\mathbf{s}_1\cdot \mathbf{s}_2 =
-3/4,1/4$ for the pseudoscalar and vector channels respectively.

%}}}

%{{{ Lattice Parameters and Correlators

{\em Lattice Parameters and Correlators --} We performed lattice
calculations of QCD with two dynamical flavours of light quark using
a Wilson-type action with anisotropy of $\xi = a_s/a_\tau = 6$, $a_s
\simeq 0.162$fm and $a_\tau^{-1} \simeq 7.35$GeV \cite{Morrin:2006tf,Oktay:2010tf}.
The other lattice parameters are listed in Table \ref{tab:params}.  We
note that the range of temperatures is from the confined phase up to
$\sim 1.7T_c$ where $T_c$ is the deconfining transition. The charm
quark is simulated with the (anisotropic) clover action and its mass
is set by matching the experimental $\eta_c$ mass at zero temperature.

\begin{table}[h]
\begin{center}
\begin{tabular}{ccccr}
\hline
 \multicolumn{1}{c}{$N_s$} & \multicolumn{1}{c}{$N_\tau$} & \multicolumn{1}{c}{$T$(MeV)} &
\multicolumn{1}{c}{$T/T_c$} & \multicolumn{1}{c}{$N_{\rm cfg}$} \\
\hline 
12 & 80 &  90 & 0.42 & 250  \\      
12 & 32 & 230 & 1.05 & 1000 \\ 
12 & 28 & 263 & 1.20 & 1000  \\ 
12 & 24 & 306 & 1.40 &  500  \\ 
12 & 20 & 368 & 1.68 & 1000  \\ 
\hline
\end{tabular}
\caption{Lattice parameters used, including spatial and temporal
  dimension, $N_s$ and $N_\tau$, temperature, and number of configurations, $N_{\rm cfg}$.}
\label{tab:params}
\end{center}
\end{table}

%}}}

%{{{ Results

\begin{figure}[t]
 \centerline{\includegraphics*[width=8.5cm]{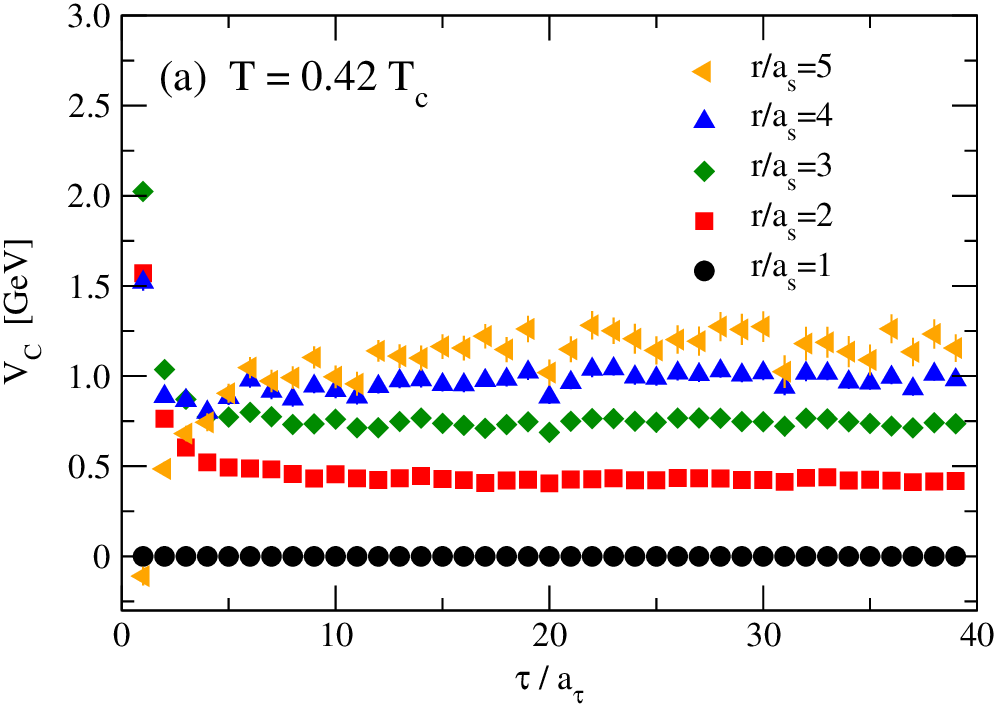}}
 \centerline{\includegraphics*[width=8.5cm]{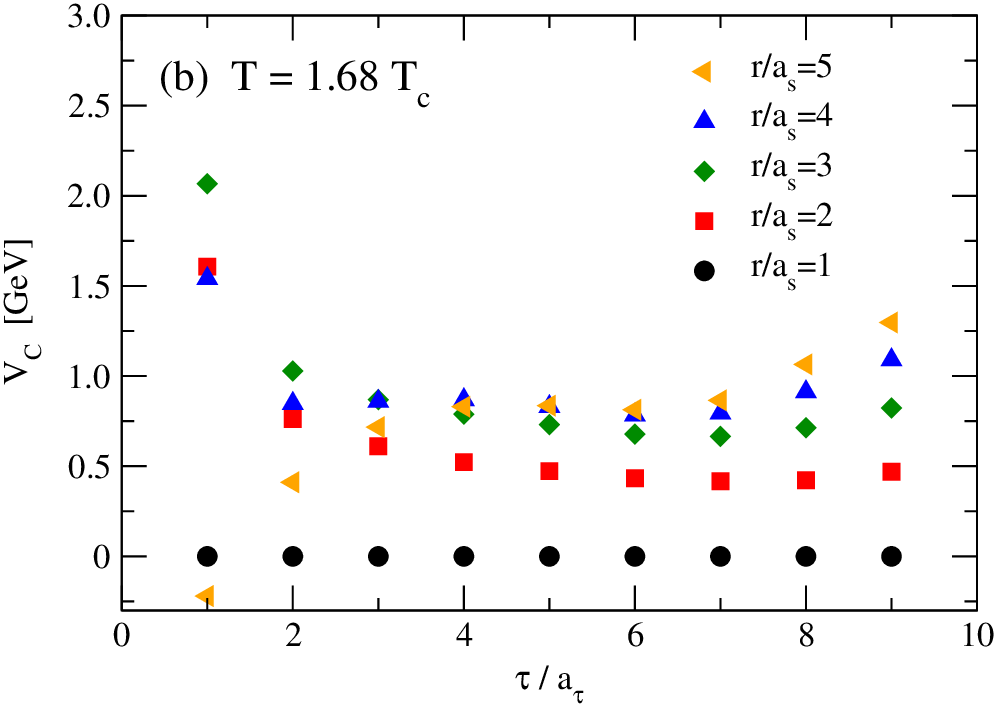}}
\caption{
The results for the central potential $V_C(r)$ obtained from
Eq.~(\ref{eq:V}) for (a) $T = 0.42 T_c$ and (b) $1.68 T_c$.
The horizontal axis is the Euclidean time, $\tau$, appearing in 
  Eq.~(\ref{eq:V})}.
 \label{fig:pot-t}
\end{figure}

{\em Results --} We now apply Eq.~(\ref{eq:V}) to obtain the potential,
$V_\Gamma(r)$, for the temperatures listed in Table \ref{tab:params}
for the vector and pseudoscalar channels separately.  In
Eq.~(\ref{eq:V}), standard symmetric lattice finite differences are used
for the spatial and temporal derivatives.  Figure \ref{fig:pot-t}
shows the central potential obtained for $T/T_c = 0.42$
and $1.68$ as a function of the time, $\tau$, appearing in
Eq.~\eqref{eq:V}. For each $\tau$ value in Fig.\ref{fig:pot-t}, we have
vertically shifted the data points so that $V_C(r/a_s=1) = 0$.

As can be seen, there is a good plateau where the potential $V_C(r)$
is stable.  The lack of a plateau at small times is presumably due to
lattice artefacts caused by contact terms at the source. The upward
trend of data points at large times and high temperature corresponds
to time values close to the centre of the lattice which are
contaiminated by backward moving states. We have confirmed this
interpretation by successfully modelling the effects of these backward
moving states.

\begin{figure}[th!]
 \centerline{\includegraphics*[width=8.5cm]{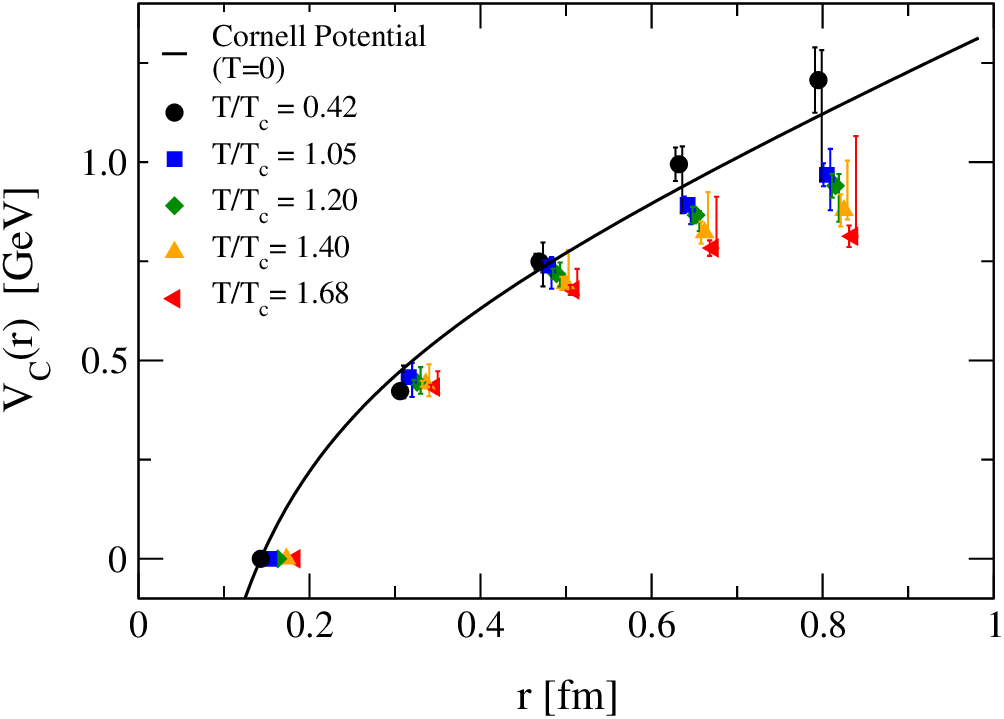}}
 \caption{
   Spin independent (i.e. central) potential, $V_C(r)$, for the
   temperatures in Table \ref{tab:params} obtained from
   eq.(\ref{eq:pot-cs}).
   The data points have been shifted horizontally for clarity.
   The solid curve is a fit to the Cornell
   potential \cite{Eichten:1979ms} (see text).}
 \label{fig:potc}
\end{figure}

\begin{figure}[th!]
 \centerline{\includegraphics*[width=8.5cm]{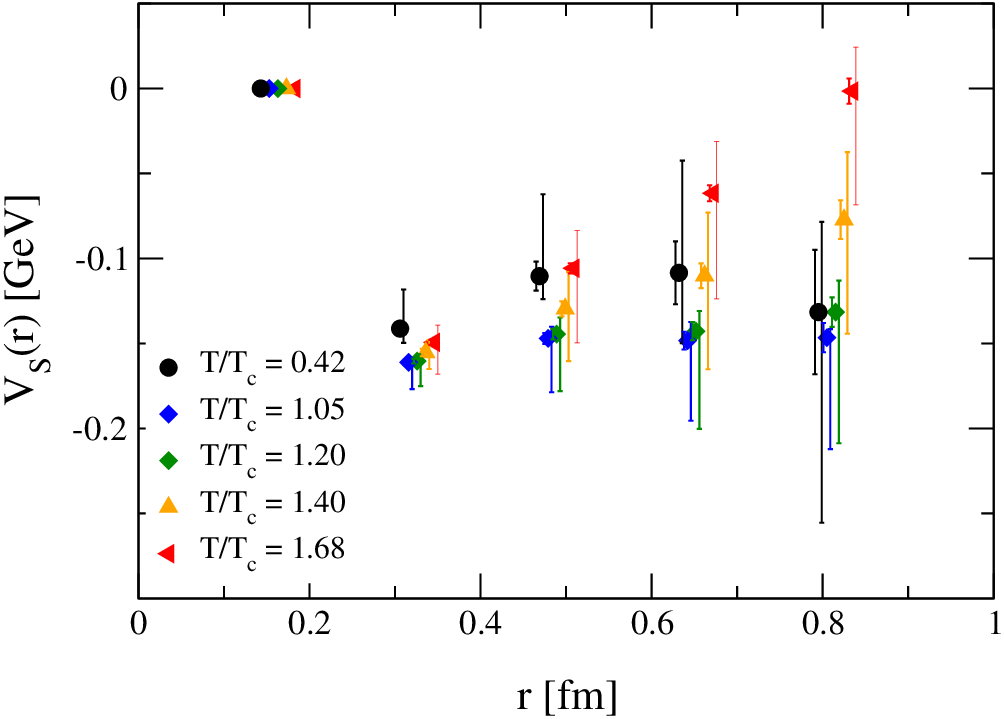}}
 \caption{Spin dependent potential, $V_S(r)$, for the
   temperatures in Table \ref{tab:params} obtained from
   eq.(\ref{eq:pot-cs}).
   The data points have been shifted horizontally for clarity.}
 \label{fig:pots}
\end{figure}

The central values for the potentials are obtained from $\tau =
6,7,7,7$ and $24$ for $N_T=20,24,28,32$ and $80$ respectively.  The
resulting $V_C$ and $V_S$ are shown in Figs. \ref{fig:potc} and
\ref{fig:pots}. The left-hand error bars are statistical, and the
systematic uncertainty of choosing different values of $\tau$ to
define the potentials are depicted in the right-hand error bar.  In
Fig.\ref{fig:potc} we include the Cornell potential, $V(r) =
-\frac{\kappa}{r} + \frac{r}{a^2} + V_0$ with $\kappa = 0.52$ and $a =
2.34$ GeV$^{-1}$ \cite{Eichten:1979ms}, as a point of reference.


From Fig.\ref{fig:potc} we see a clear temperature dependence. In the
confined phase, $T=0.42 T_c$, we see evidence of a linearly rising
potential in agreement with the Cornell potential. As the temperature
increases beyond $T_c$, the potential flattens for large distances, in
agreement with expectations of a deconfined phase.  The spin dependent
potential is plotted in Fig.\ref{fig:pots} and shows a repulsive core.

We now compare our results with those using static quarks.  There are
two general approaches to extract the interquark potential between
static (infinitely heavy) quarks, both of which have limitations.
The first calculates the free energy of a static quark pair as a
function of their separation via various correlators of Polyakov loops
\cite{static-pot,Kaczmarek:2005ui,Bazavov:2012fk}. However,
the potentials thus derived suffer from either gauge dependence (in
the case of the ``singlet'' channel), or do not reduce to the correct
Debye screened potential in the perturbative limit (in the case of the
``averaged'' channel) \cite{philipsen,Bazavov:2012fk}.
The second uses Wilson loops or correlators of Wilson lines
\cite{rothkopf,Bazavov:2012fk,Burnier:2012az} and requires there to be
good ground state dominance. However, this creates tension because the
temporal extent of the lattice, $N_\tau \sim 1/T$ is necessarily small
at high temperature. As a result, precision results are difficult to
obtain.

The method discussed here is gauge invariant by construction and
produces results with reasonable systematics. Furthermore it
calculates the potential between quarks with masses tuned to the
physical charm.

\begin{figure}[th!]
 \centerline{\includegraphics*[width=8.5cm]{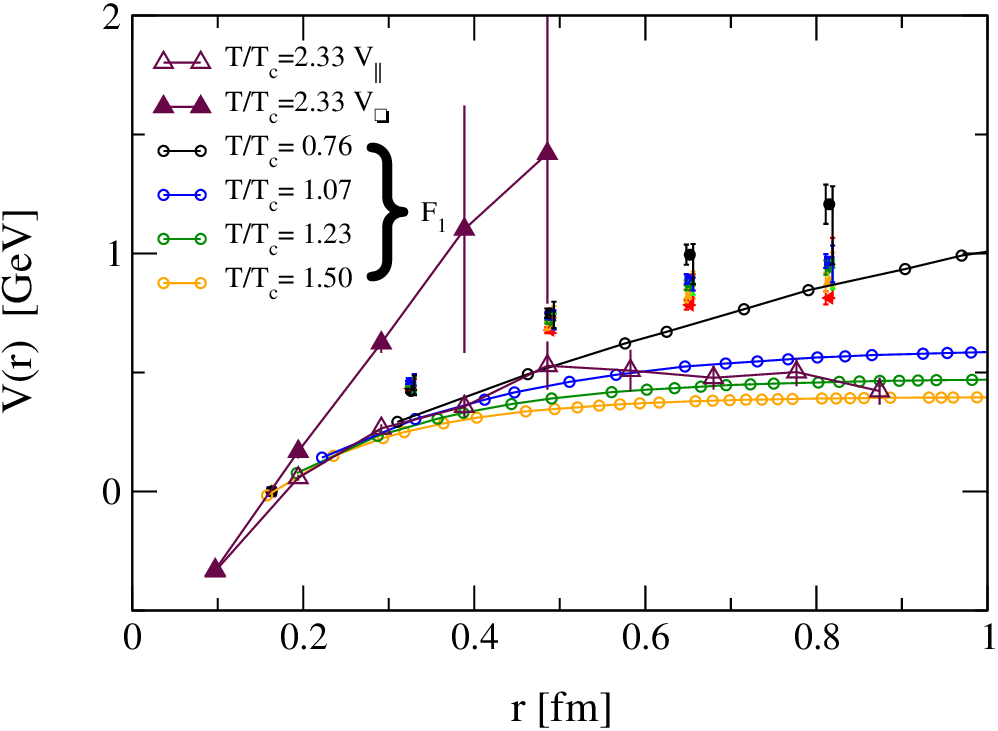}}
 \caption{ Comparison of $V_C$ from this work with the singlet free
   energy calculation, $F_1$, from \cite{Kaczmarek:2005ui} and
   the Wilson loop, $V_\square$, and Wilson line correlator, 
     $V_{||}$, from \cite{Burnier:2012az}.  The error bars of the
   free energy data are smaller than the symbols.  The data from this
   work follows the legend in Fig. \ref{fig:potc}.  }
 \label{fig:pot-static}
\end{figure}

In Fig.\ref{fig:pot-static} we compare our results from
Fig.\ref{fig:potc} with those obtained from static quark
calculations -- the singlet free energy \cite{Kaczmarek:2005ui} and the
Wilson loop and line \cite{Burnier:2012az}.  We note a clear discrepancy
between the our results and those obtained from static quarks.

%}}}

%{{{ Conclusions and Outlook

{\em Conclusions --} There is a significant body of theoretical work
studying the interquark potential at non-zero temperature using model,
perturbative and lattice (non-perturbative) approaches.  The work
outlined here uses a lattice simulation of QCD with two light
dynamical flavours on an anisotropic lattice. We determine the
charmonium potential at a variety of temperatures using relativistic
quarks tuned to the physical charm quark mass. This improves upon
earlier lattice simulations performed in the static limit.  It thus
represents the first ab initio calculation of the charmonium
potential of QCD at finite temperature.

The method we use is based on the HAL QCD ``time-dependent'' approach
which obtains the real part of the potential from correlators of
point-split operators \cite{halqcd}. This allows the extraction of the
potential without the need to first define the NBS wavefunctions by
fitting the large time behaviour of the correlation functions.

Our determination of the potential shows a linearly rising potential
for $T < T_c$ and a clear temperature dependent flattening of the
potential for $T > T_c$. We demonstrate a significant deviation
between our results and those obtained using static quarks via either
the free energy or Wilson loops/lines.

This work adds to previous charmonium studies performed by our
collaboration with the same lattice parameters
\cite{Aarts:2007pk,Oktay:2010tf} and
our earlier work on the potential using the HAL QCD wavefunction method
\cite{Allton:2012ki}.

In forthcoming work we will simulate on significantly larger lattices
with 2+1 light quark flavours. We also hope to extend our work
to the potential between heavier quarks using the NRQCD approach
\cite{Aarts:2010ek,Aarts:2011sm,Aarts:2012ka}.

%}}}

%{{{ Acknowledgements

\smallskip

We acknowledge the support and infrastructure provided by the Trinity
Centre for High Performance Computing and the IITAC project funded by
the HEA under the Program for Research in Third Level Institutes
(PRTLI) co-funded by the Irish Government and the European Union.  The
calculations have been carried out using CHROMA
\cite{Edwards:2004sx}. The work of CA and WE is carried as part of the
UKQCD collaboration and the DiRAC Facility jointly funded by STFC, the
Large Facilities Capital Fund of BIS and Swansea University. WE and CA
are supported by STFC. CRA thanks the Galileo Galilei Institute for
Theoretical Physics for hospitality and the INFN for support during
the writing up of this work. We are very grateful to Gert Aarts, Sinya
Aoki, Robert Edwards, Tetsuo Hatsuda, Balint Jo\'o and Alexander
Rothkopf for useful discussions.

%}}}

%{{{ bibliography

%}}}

\end{document}